\documentclass[a4paper,12pt]{article}

\usepackage{amsmath,amssymb,amscd,amsthm,mathtools,array,empheq}
\usepackage[utf8]{inputenc}
\usepackage{lmodern,newtxtext,newtxmath}
\usepackage[a4paper,tmargin=3truecm,bmargin=3truecm,hmargin=1.6truecm]{geometry}
\usepackage[unicode,colorlinks=true, urlcolor=blue, citecolor=blue, linkcolor=blue]{hyperref}
\usepackage{graphicx}
\usepackage{color}
\usepackage[dvipsnames]{xcolor}
\usepackage[english]{babel}
\usepackage{bm}
\usepackage{bbm}
\usepackage{setspace}
\usepackage{cite}
\usepackage{etoolbox}
\apptocmd{\thebibliography}{\setlength{\itemsep}{0pt}}{}{}
\usepackage[shortlabels]{enumitem}
\numberwithin{equation}{section}
\usepackage[normalem]{ulem}

\newif\ifdev
\devfalse

\newtheoremstyle{Definition}
{\topsep}% Space above
{\topsep}% Space below
{\itshape}% Body font
{}% Indent amount
{\bfseries}% Theorem head font
{\newline}% Punctuation after theorem head
{1em}% Space after theorem head
{\thmname{#1}~\thmnumber{#2}\thmnote{\ -\  #3}}% Theorem head spec

\theoremstyle{Definition}

\usepackage{tikz}
\usetikzlibrary{shapes, arrows, arrows.meta, matrix, decorations.markings, calc, babel, quotes, angles, cd}
\makeatletter \g@addto@macro\@floatboxreset\centering \makeatother

\newcommand{\cM}{\mathcal{M}}

\newcommand{\cU}{\mathcal{U}}

\newcommand{\bg}{\boldsymbol{g}}

\newcommand{\bq}{\boldsymbol{q}}
\newcommand{\bp}{{\boldsymbol{p}}}
\newcommand{\br}{{\boldsymbol{r}}}

\newcommand{\bx}{{\boldsymbol{x}}}

\newcommand{\lieg}{\mathfrak{g}}

\usepackage{scalerel}
\let\savewidetilde\widetilde
\def\widetilde#1{%
 \ThisStyle{\savewidetilde{\phantom{\SavedStyle#1}}%
  \setbox0=\hbox{$\SavedStyle#1$}\kern-\wd0#1}}

\title{Statistical Physics of Planar Carroll Systems}
 
\author{Fei Huang\footnote{huangf93@mail2.sysu.edu.cn}, Loïc Marsot\footnote{Corresponding author: marsot3@mail.sysu.edu.cn}\\[1.em]
{\normalsize School of Science, Shenzhen campus of Sun Yat-sen University,} \\
{\normalsize Shenzhen, Guangdong 518107, P.R. China}
} %authors

\date{{\footnotesize (\today)}}

\begin{document}

\maketitle

\begin{abstract}
In this article, we define and study the statistical physics of planar Carrollian systems. While it has been shown recently that, for general dimensions, the Carroll limit of Poincaré statistical physics typically does not converge, we show that thanks to the central extensions of the Carroll algebra in the plane, and by considering systems with angular momentum, there exists a well defined notion of planar Carrollian statistical physics.

Using Souriau's geometric thermodynamics, we compute the partition function for particles on a uniformly rotating disc, and show that rotation is inevitable for thermal equilibrium of planar Carroll systems, with one of the central charges determining the direction of rotation. We derive thermodynamic quantities in particular entropy, which scales logarithmically with the disc area, and pressure, which follows the two-dimensional ideal gas law. Though all results are obtained from symmetry considerations, we also derive the corresponding effective Hamiltonian.
\end{abstract}

\section{Introduction}

In 1965, Lévy-Leblond discovered the Carroll group as an ultra-relativistic contraction of the Poincaré group \cite{LvyLeblond1965UneNL}, where $c \to 0$ but $v / c \ll 1$. Despite the fact that the Carroll group is less known than the Poincaré and Galilei groups, it is still one of the possible kinematical groups, as was shown by Bacry and Lévy-Leblond at the end of the 1960s \cite{Bacry:1968zf}. For a long time, this group was regarded as little more than a mathematical curiosity, since the causal cone closes up, which in particular implies that free particles obeying Carroll symmetry cannot move \cite{LvyLeblond1965UneNL,SenGupta:1966qer,Bacry:1968zf,Duval:2014uoa,Bergshoeff:2014jla}.

Nevertheless, physical applications began to be considered in General Relativity about a decade ago. First, it was recognized that the famous Bondi-Metzner-Sachs (BMS) group (\textit{i.e.} the group of symmetries at null infinity in asymptotically flat spacetimes) \cite{Bondi:1962px,PhysRev.128.2851}, is a conformal extension of the Carroll group (on an $S^2 \times \mathbb{R}$ topology) \cite{Duval:BMS05,Duval:BMS08}. This prompted the use of Carroll symmetry in flat space holography, see \cite{Ciambelli:2018wre,Campoleoni:2018ltl,PhysRevLett.129.071602,Donnay:2022wvx,Saha:2023hsl,Bagchi_2022,Bagchi:2022owq,Bagchi:2023fbj}. Moreover, ``Carroll physics'' is found to be relevant on the horizon of black holes \cite{Penna:2018gfx,Donnay:2019jiz,Marsot:2022qkx,Redondo-Yuste:2022czg,Freidel:2022bai,Freidel:2022vjq,Freidel:2024emv} (as an application of null hypersurfaces), and for gravitational waves, in particular on the memory effect \cite{Duval:2017els,Zhang2017TheME,Zhang2017SG}. Additionally, Carroll symmetry appears in string theory \cite{Bagchi:2023cfp,Blair:2023noj,Gomis:2023eav,Bagchi:2024rje,Banerjee:2025bkg} and also in condensed matter physics, which might provide the most accessible experimental realizations of Carroll systems, in particular in some specific configuration of graphene \cite{Bagchi:2022eui}, and for so-called fractons \cite{Bidussi:2021nmp,Marsot:2022imf,Figueroa-OFarrill:2023vbj}.

Nearly all these applications, including in condensed matter, occur ``in the plane'', meaning on $2 + 1$ dimensional spacetime. In fact, as argued in \cite{Morand:2018tke,PhysRevD.100.046010}, any null hypersurface embedded in a Lorentzian manifold is a Carroll geometry \cite{Henneaux:1979vn,Duval:1990cele,Dautcourt:1997hb,Duval:2014uoa}. This dimension of application makes the study of planar Carroll systems important, especially as recalled below, the Carroll symmetries in the plane are much richer than in other dimensions\footnote{Except perhaps in $1+1$ dimensions, but in this case the Carroll group is isomorphic to the Galilei group.}.

Given generators $J_i, P_i, K_i, M$ respectively representing rotations, spatial translations, boosts, and time translation, the non-trivial commutators of the Carroll algebra in $3 + 1$ dimensions are:
\begin{equation}
[J_i, J_j] = \epsilon_{ijk} J_k, \quad
[J_i, P_j] = \epsilon_{ijk} P_k, \quad
[J_i, K_j] = \epsilon_{ijk} K_k, \quad
[K_i, P_j] = M \delta_{ij}
\end{equation}
The distinctive feature of the Carroll group is that boosts affect time coordinate, in contrast to the Galilei group where they affect the spatial coordinates instead. 

Crucially, in $2 + 1$ dimensions, the Carroll algebra turns out to be richer since it admits two non-trivial central extensions \cite{deAzcarraga:1997mdw,Ngendakumana:2013hza} generated by $A_1$ and $A_2$:
\begin{equation}
\begin{split}   
[J_3, P_i] & = \epsilon_{ij} P_j, \quad
[J_3, K_i]   = \epsilon_{ij} K_j, \quad
[K_i, P_j]   = M \delta_{ij},     \\
[P_i, P_j] & = \epsilon_{ij} A_1, \quad
[K_i, K_j]   = \epsilon_{ij} A_2
\end{split}
\end{equation}

Central extensions are essential in physics. The notion of mass in Galilean relativity exists because the Galilean group has a non-trivial central extension \cite{Bargmann54,levy-leblond1971}, mass being the associated conserved quantity. This is not the only example of central extension. For instance, in $2 + 1$ dimensions the Galilei group admits two central extensions \cite{levy-leblond1971,ballesteros_1992_moyal,Lukierski:1996br,duval_2000_the,Duval:2001hu}, labeled by mass and an exotic parameter. This doubly-extended Galilei group is shown to be realized in quantum mechanics on the non-commutative plane, as quantization of this planar system coupled to a magnetic field \cite{Duval:2001hu,duval_2000_the} recovers the Laughlin description \cite{PhysRevLett.50.1395} of the ground states of the Fractional Quantum Hall Effect. In the matter at hand, the two central extensions of the Carroll group notably allows for motion \cite{Marsot21}, which make them interesting to study. 

To provide experimental observables related to the Carroll symmetry, we study the statistical physics and thermodynamic quantities of planar Carroll systems. The central extensions not only allow for motion, but also enter the partition function and are crucial for a well-defined statistical physics. While the partition function of Carroll systems has been computed in general dimensions as the zero speed of light limit of Poincaré ones \cite{deBoer:2023fnj} and found to diverge, we show that convergent quantities emerge when considering rotating systems together with the central extensions.

The organization of this paper is as follows. In Section \ref{sec:statistical_physics} we review two equivalent methods to calculate the partition function, one relying on Hamiltonian and the other on conserved quantities induced from a symmetry group. Subsequently, using the second approach, in Section \ref{sec:Carroll_rotating} we investigate the statistical physics of planar Carroll particles restricted on a uniformly rotating disc, where we show that thermodynamic quantities converge. We also find the associated statistical distributions, gas law, and effective Hamiltonian. Final conclusions are given in Section \ref{sec:conclusions}.

\section{Statistical Physics of a Dynamical Group}
\label{sec:statistical_physics}

In statistical physics, we are concerned with calculating partition functions $Z$, and then deriving macroscopic observables from these functions. For example, for a canonical ensemble, which is a system with fixed volume and particle number in thermal equilibrium with its environment, the partition function of one particle is given by
%\begin{equation}
%Z =  \sum_i e^{- \beta E_i}
%\end{equation}
%where $\beta = 1 / k_B T$, with $T$ the temperature, and $E_i$ is the total energy of the system in the microstate $i$.
%
%The probability $\rho_i$ that the system is in the microstate $i$ is $\rho_i = \frac{e^{-\beta E_i}}{Z}$, and the average energy of the system is then 
%\begin{equation}
%U = \sum_i \rho_i E_i
%\end{equation}
%together with its entropy
%\begin{equation}
%S = \frac{U}{T} + k_B \ln Z
%\end{equation}
%
%For a canonical ensemble in classical mechanics, the partition function is defined as
\begin{equation}
\label{oneparticle_partition}
Z = \frac{1}{h^3} \int_\cU e^{-\beta H(\bq, \bp)} d^3 \bq d^3 \bp
\end{equation}
where $\beta = 1 / k_B T$, with $T$ the temperature, $H$ is the Hamiltonian of the particle, and $\cU$ is the (symplectic) phase space.

%For $N$ identical particles with negligible interaction energy, we can factorize the total partition function as the product of the individual partition functions. However one needs to be careful if particles are identical and we cannot distinguish them, as one would need to divide by the number of permutations of the $N$ identical particles to correctly count the microstates,
%\begin{equation}
%Z = \frac{1}{N! h^{3N}} \int \prod_{i=1}^N e^{-\beta H(\bq_i, \bp_i)} d^3 \bq_i d^3 \bp_i
%\end{equation}

The partition function is particularly useful to calculate macroscopic observables, for example the average energy $U$ is obtained as
\begin{equation}
U = - \frac{\partial \ln Z}{\partial \beta}
\end{equation}
and the entropy as
\begin{equation}
S = k_B \left(\ln Z + \beta U\right)
\end{equation}

The partition function tells us how the microstates are partitioned, and in the standard presentation of statistical physics displayed above, we see the partition function depends on the pair of conjugate variables $(\beta, H)$ of the temperature and energy of the components, where the energy is a conserved quantity of the system. A natural question is then: if (the phase space of) a dynamical system is invariant under a group $G$, can we extend this formulation to include all the conserved quantities generated by $G$ (or, in general, the central extension of $G$)?

To illustrate this, consider the example of a gas in a centrifuge. There are two ways to find the partition function of a particle in this gas. The most common way of doing so is to consider that the centrifuge induces a centrifugal potential on each particle, and hence contributes to the energy, see \textit{e.g.} \cite{LandauLSP}. For an angular velocity $\omega$ and a distance $r$ from the particle to the rotation axis, this potential is $- m \omega^2 r^2/2$, and the conserved energy thus becomes $\frac{\bp^2}{2m} - \frac{m}{2} \omega^2 \br^2$.

The one-particle partition function is then
\begin{equation}
Z = \int_\cU e^{-\beta \left(\frac{\bp^2}{2m} - \frac{m}{2} \omega^2 \br^2\right)} d^3 \bp d^3\bq
\end{equation}

The other point of view amounts to consider, instead of the centrifugal potential, that the centrifuge induces a non-zero angular momentum to the system, and couple the angular momentum with its conjugate variable in the partition function alongside the purely kinematic energy and its conjugate variable. The two views are equivalent, as can be seen by recasting the above expression, using the change of variables $p_1 \mapsto p_1 - m \omega x_2$ and $p_2 \mapsto p_2 + m \omega x_1$ \footnote{This transformation amounts to switch between an inertial and a rotating frame.}, into the form,
\begin{equation}
Z = \int_\cU e^{\beta \omega (x_1 p_2 - x_2 p_1) - \beta \frac{\bp^2}{2m}} d^3 \bp d^3\bq
\end{equation}

%Hence, the relevant quantity for the computation of the partition function is
%\begin{equation}
%\mu \cdot \psi(x) = \alpha (r_1 p_2 - r_2 p_1) - \frac{\beta}{2m}\left(\bp^2 - m^2 \omega^2 \left(r_1^2 + r_2^2\right)\right)
%\end{equation}
%where $\alpha$ is some parameter akin to $\beta$ but for rotations instead of temperature.

Clearly $x_1 p_2 - x_2 p_1$ is the angular momentum of the system in the $(Oz)$ direction, and $\beta \omega$, which has dimension of the inverse of an angular momentum is the conjugate variable. The structure of the argument of the exponential is hence $\alpha L - \beta H$, for $\alpha = \omega \beta$.

In general, it has been argued \cite{Souriau70} that the argument of the exponential is $\mu(\bp, \bq) \cdot \xi$ where $\mu : \cU \to \lieg^*$ is what is known as the moment map\footnote{Although \emph{momentum map} would be more appropriate in English \cite{MarsdenR94}.} of the particle (\textit{i.e.} the set of its conserved quantities), and $\xi \in \lieg$ is the set of conjugate variables. The dot is then the natural pairing of the Lie algebra $\lieg$ of $G$ and its dual $\lieg^*$, given some sign conventions.

In the rotating example, the component corresponding to time translation symmetry is the temperature $\beta$, while the component corresponding to rotations corresponds to some ``angular momentum temperature'', $\alpha$ in the example of the previous paragraph. Explicitly, $\xi = (\alpha, \beta)$. 

In particular, when we only consider time translations, we obtain the average energy as $U = - \frac{\partial \ln Z}{\partial \beta}$. In general, \cite{Souriau70} one obtains the set of average quantities as $M = \frac{\partial \ln Z}{\partial \xi} \in \lieg^*$. In the rotating gas example, since we have angular momentum, we also have the average total angular momentum as $L = \frac{\partial \ln Z}{\partial \alpha}$. As for the entropy, it becomes in general
\begin{equation}
\label{general_entropy}
S = k_B \left(\ln Z - M \cdot \xi\right)
\end{equation}
Differentiating this relation, one gets the first law of thermodynamics,
\begin{equation}
\label{general_first_law}
dS = - k_B \xi \cdot dM
\end{equation}
In our example of a rotating gas, 
\begin{equation}
S = k_B \left(\ln Z + \beta U - \alpha L\right)
\end{equation}
and
\begin{equation}
dU = T dS + \omega dL
\end{equation}
in accordance with e.g. \cite{LandauLSP}.

Note that central extensions play an important role in this description of statistical physics as well. Consider a Galilean system whose relevant conserved quantities are the energy $H = p^2/2m$, and the mass $m$, which is the conserved quantity associated to the central extension algebra generator. Write the conjugate quantities as $\xi = (\beta, \gamma)$, and the average total quantities as $M = (U, \cM)$. The partition function for one particle is $Z_1 = \int_\cU e^{-\beta H + \gamma m} d\lambda$, and for $N$ particles, using Stirling's approximation, $z = \ln Z_1^N/N! \approx N \left(\gamma m + \ln \left[\left(\frac{2 \pi m}{\beta}\right)^{3/2} \frac{V}{N} \right] + 1 \right)$, which is such that the total mass is
\begin{equation}
\cM = \frac{\partial z}{\partial \gamma} = N m
\end{equation}
Writing the first law, considering the number of particles $N$ variables, and defining $\mu_0 = \gamma m / \beta$, we find
\begin{equation}
d U = T dS - \mu_0 dN
\end{equation}
meaning that the Galilean central extension is related to the chemical potential.

One advantage of this formalism, whose geometry is much more detailed in \cite{Souriau70}, is that all the thermodynamical objects are intrinsically invariant (such as the entropy) or covariant with respect to the transformations of the group $G$.

This formalization is not usually needed, since we can usually calculate the partition function from the particle's Hamiltonian with \eqref{oneparticle_partition}. However, if one wishes to study the statistics of some dynamical system based on some exotic symmetry group $G$, where the form of Hamiltonians is not clear, this alternative point of view provides interesting, and more certain, insights.

For more details about this ``Geometric thermodynamics'' formulation, see \textit{e.g.} the more recent \cite{Marle16,Barbaresco16,FreST26}.

\section{Carroll Particles on a Uniformly Rotating Disc}
\label{sec:Carroll_rotating}

\subsection{Theoretical Framework}
\label{sec_framework}

Define abstractly the set of generators of the extended Carroll algebra $\xi = (\alpha, \Upsilon, \bm{\gamma}, \beta, \zeta_1, \zeta_2)$, and the set of conserved quantities $\mu=(l, \bm{g}, \bm{P}, m, q_1, q_2)$. Since $\xi$ and $\mu$ belong to dual spaces, one can define their natural pairing
\begin{equation}
\mu \cdot \xi = \alpha l + \Upsilon \cdot \bm{g} + \bm{\gamma} \cdot \bm{P} - \beta m + \zeta_1 q_1 + \zeta_2 q_2
\end{equation}
where the signs are chosen for later convenience, for example to define the energy, or the mass, as positive.

For a free planar particle whose kinetics is governed by the extended Carroll group, the conserved quantities are obtained from, respectively, rotations, boosts, spatial translations, and the three generators in the center of the group, and are calculated to be \cite{Marsot21}
\begin{subequations}
\begin{align}
\label{conserved_quantity_rotations}
l & = \bm{p} \times \bm{x} + q_1 \bm{x}^2 - \frac{q_2}{m^2} \bm{p}^2 \\
\label{conserved_quantity_boost}
\bm{g} & = m \bm{x} + \frac{2q_2}{m} \epsilon \bm{p} \\
\bm{P} & = \bm{p} + 2 q_1 \epsilon \bm{x} \\
m   \\
q_1 \\ 
q_2
\end{align}
\end{subequations}
Note that $\bm{p} = m \bm{v}$ is the momentum coordinate, to be distinguished from $\bm{P}$ which denotes the conserved momentum.

In Carroll theories, calling the parameter dual to the time translation generator a mass or an energy is merely a choice of units. However, sometimes the behavior of this parameter looks strikingly similar to the mass in Galilean systems\footnote{In fact, the Carroll group is a subgroup of the Bargmann group, which is the central extension of the Galilei group, and the Carroll time translations correspond to the central extension generator, whose dual quantity in Galilean systems is the mass.}. However, some other times, in particular in thermodynamics, the corresponding parameter, here typically the ensemble average behaves like an energy. For the sake of comparison with existing theories, we choose to write the physical parameter dual to Carroll time translations as a mass $m$, but later we will call its thermodynamical ensemble average $U$, reminiscent of the notation for the energy in Galilean thermodynamics.
 
In the following subsections, we will consider Carroll particles with angular momentum, mass, $q_1$, $q_2$, and the corresponding conjugate quantities, but with vanishing $\Upsilon$ and $\bm{\gamma}$ such that we can ignore the particles' $\bg$ and $\bm{P}$ on a uniformly rotating disc. Their physical state will be described by $\mu=(l, \bg, \bm{P}, m, q_1, q_2)$ and $\xi = (\alpha, 0, 0, \beta, \zeta_1, \zeta_2)$. In this case, the pairing takes the form
\begin{equation}
\label{pairing_model}
\mu \cdot \xi = \alpha \left( \bm{p} \times \bm{x} + q_1 \bm{x}^2 - \frac{q_2}{m^2} \bm{p}^2 \right) 
- \beta m + \zeta_1 q_1 + \zeta_2 q_2
\end{equation}

As argued in Section \ref{sec:statistical_physics}, the one-particle partition function of this system is given by
\begin{equation}
Z = \int_{\cU} e^{\mu \cdot \xi} d^2 \bm{p} d^2 \bm{x}
\end{equation}
such that the distribution $\gamma_\xi$ of observables over phase spaces $h \in C^\infty(\cU)$ for a given $\xi$ is given by\cite{Souriau70},
\begin{equation}
\label{distribution_phase_space}
\gamma_\xi(h) = \frac{1}{Z} \int_\cU h(\bp, \bx) e^{\mu\cdot \xi} d^2\bp d^2 \bx
\end{equation}
In particular, the average of the conserved quantities $\mu = (l, m, q_1, q_2)$ give the ensemble averages, which are equivalently calculated as,
\begin{equation}
\label{ensemble_average_def}
L   = \left.\frac{\partial \ln Z}{\partial \alpha}\right|_{\beta,\zeta_1,\zeta_2},   \quad 
U   = - \left.\frac{\partial \ln Z}{\partial \beta}\right|_{\alpha,\zeta_1,\zeta_2},  \quad 
Q_1 = \left.\frac{\partial \ln Z}{\partial \zeta_1}\right|_{\alpha,\beta,\zeta_2},  \quad
Q_2 = \left.\frac{\partial \ln Z}{\partial \zeta_2}\right|_{\alpha,\beta,\zeta_1}
\end{equation}
The entropy \eqref{general_entropy} is then
\begin{equation}
\label{entropy_carroll_def}
S = k_B \left(\ln Z - L \alpha + U \beta - Q_1 \zeta_1 - Q_2 \zeta_2 \right)
\end{equation}
and the first law of Carroll thermodynamics \eqref{general_first_law},
\begin{subequations}
\begin{equation}
dU = T dS + \omega dL + \Phi_1 dQ_1 + \Phi_2 dQ_2 
\end{equation}
\begin{equation}
\beta  = \frac{1}{k_B T},        \quad 
\omega = \frac{\alpha}{\beta},   \quad 
\Phi_1 = \frac{\zeta_1}{\beta},  \quad 
\Phi_2 = \frac{\zeta_2}{\beta}
\end{equation}
\end{subequations}
where $\beta$ is the inverse temperature as in the Galilean example, $\omega$ is the angular velocity dual to the angular momentum $L$, and $(\Phi_1, \Phi_2)$ are the potentials dual to the central charges $(Q_1, Q_2)$. While we provide the common physical interpretations of $\beta$ and $\omega$, by matching the formalism with the known Galilei thermodynamics, the interpretation of $(\Phi_1, \Phi_2)$ and the corresponding charges is not given. The framework given in this article is mathematical in nature, and physical interpretation can only be given once we observe a physical realization of this model.

There is a missing term to the first law, $- P dA$, representing the work done by area changes against pressure. Being an external contribution, it needs to be added by hand. Hence, the first law is generalized to
\begin{equation}
\label{first_law}
dU = T dS + \omega dL + \Phi_1 dQ_1 + \Phi_2 dQ_2 - P dA
\end{equation}

\subsection{Partition Function}

Consider a two-dimensional disc of radius $R$. Rotational symmetry makes polar coordinates more convenient than Cartesian coordinates. We therefore perform the following coordinate transformation which leaves the phase space volume form unchanged
\begin{equation}
\begin{cases}
x_1 = r \cos{\theta} \\
x_2 = r \sin{\theta} \\
p_1 = p_r \cos{\theta} - \frac{p_{\theta}}{r} \sin{\theta} \\  
p_2 = p_r \sin{\theta} + \frac{p_{\theta}}{r} \cos{\theta} 
\end{cases}
\end{equation}
such that
\begin{equation}
\begin{cases}
x_1 p_2 - x_2 p_1 = p_{\theta} \\
p_1^2 + p_2^2 = p_r^2 + \frac{p_{\theta}^2}{r^2}
\end{cases}
\end{equation}

The pairing \eqref{pairing_model} now becomes
\begin{equation}
\mu \cdot \xi = 
-\frac{\alpha q_2}{m^2} p_r^2 
- \frac{\alpha q_2}{m^2r^2} p_{\theta}^2 
- \alpha p_{\theta} 
+ \alpha q_1 r^2 
- \beta m + \zeta_1 q_1 + \zeta_2 q_2
\end{equation}
Consequently, we can compute the partition function
\begin{equation}
\label{partition_function_computed}
\begin{split}
Z & = e^{- \beta m + \zeta_1 q_1 + \zeta_2 q_2} \, 
\int_0^{2\pi} d\theta \,
\int_{-\infty}^{+\infty} e^{-\frac{\alpha q_2}{m^2} p_r^2} dp_r \, 
\int_0^R e^{\alpha q_1  r^2} \, \left( 
\int_{-\infty}^{+\infty} e^{- \frac{\alpha q_2}{m^2 r^2} p_{\theta}^2 - \alpha p_{\theta}} dp_{\theta} \right) dr \\
& = e^{- \beta m + \zeta_1 q_1 + \zeta_2 q_2} \,
\frac{2 \pi^2 m^2}{\alpha q_2} \,
\int_0^R r e^{\frac{\widetilde{m}^2 \alpha}{4q_2} r^2} dr 
\end{split}
\end{equation}
Here, the integral over momentum space leads to the convergence condition:
\begin{equation}
\label{convergence_condition}
\alpha q_2 = \beta \omega q_2 > 0 \overset{\beta > 0}{\implies} \omega q_2 > 0  
\end{equation}

The remaining integral in \eqref{partition_function_computed} is divided into two cases,
\begin{equation}
\int_0^R r e^{\frac{\widetilde{m}^2 \alpha}{4 q_2} r^2} dr =
\begin{cases}
\frac{R^2}{2} & \text{when $\widetilde{m}^2 = 0$} \\
\frac{2q_2}{\widetilde{m}^2 \alpha} \left( e^{\frac{\widetilde{m}^2 R^2 \alpha}{4 q_2}} - 1 \right) & \text{when $\widetilde{m}^2 \ne 0$}
\end{cases}
\end{equation}
where we define $\widetilde{m}^2  = m^2 + 4 q_1 q_2$ as the effective mass squared\footnote{Note that \textit{a priori} nothing prevents this effective mass squared from being negative.}. This leads to the finite partition function:
\begin{subequations}
\label{result_partition_function}
\begin{align}
\label{partition zero effective mass}
Z_{\widetilde{m}^2 = 0} & = \frac{\pi^2 m^2 R^2}{\alpha q_2} e^{- \beta m + \zeta_1 q_1 + \zeta_2 q_2} \\
\label{partition non-zero effective mass}
Z_{\widetilde{m}^2 \ne 0} & = \frac{4 \pi^2 m^2}{\alpha^2 \widetilde{m}^2} e^{- \beta m + \zeta_1 q_1 + \zeta_2 q_2} \left( e^{\frac{\widetilde{m}^2 R^2 \alpha}{4 q_2}} - 1 \right) 
\end{align}
\end{subequations}
The partition function $Z$ is continuous at $\widetilde{m}^2 = 0$ since $\lim_{\widetilde{m}^2 \to 0} Z_{\widetilde{m}^2 \ne 0} = Z_{\widetilde{m}^2 = 0}$. Nevertheless, it is convenient to treat them separately, because the expressions simplify considerably for $\widetilde{m}^2 = 0$.

\subsection{Ensemble Averages}
\label{sec:angular momentum}

From the partition function derived in \eqref{result_partition_function} and the definitions of the ensemble averages \eqref{ensemble_average_def}, we find that the internal energy and the average central charges are simply:
\begin{equation}
U = m, \qquad Q_1 = q_1, \qquad Q_2 = q_2
\end{equation}
while the average angular momentum is given by a complicated expression:
\begin{equation}
\label{result_average_angular_momentum}
L = -\frac{2}{\alpha} \left[ 1 
- \frac{\widetilde{m}^2 R^2 \alpha}{8 q_2} 
e^{\frac{\widetilde{m}^2 R^2 \alpha}{4q_2}} \left( 
e^{\frac{\widetilde{m}^2 R^2 \alpha}{4q_2}}-1 \right)^{-1} \right] 
\end{equation}
Taking the limit $\widetilde{m}^2 \to 0$, this expression reduces to
\begin{equation}
L_{\widetilde{m}^2 = 0} = - \frac{1}{\alpha}
\end{equation}

One question of particular interest is the asymptotic behavior of these Carroll particles when the angular velocity of the disc becomes large: does their angular momentum diverge or converge? Since $\alpha = \omega \beta$, the limit $\omega \to +\infty$ is equivalent to sending $\alpha \to +\infty$ in \eqref{result_average_angular_momentum}. The average angular momentum becomes:
\begin{subequations}
\begin{equation}
\lim_{\omega \to +\infty} L = \frac{\widetilde{m}^2 R^2}{4 q_2}, \qquad 
\text{when $\widetilde{m}^2 > 0$}
\end{equation}
\begin{equation}
\lim_{\omega \to +\infty} L = 0, \qquad \qquad 
\text{when $\widetilde{m}^2 < 0$}
\end{equation}
\end{subequations}
Note that the limit can be taken for $\omega \to -\infty$ as well, but due to the equilibrium condition \eqref{convergence_condition}, the product $\omega q_2$ has to be positive, thus leading to the same result.

Somewhat counter-intuitively, the average angular momentum diverges for $\omega \to 0$ but converges to a constant (depending on the sign of $\widetilde{m}$) in the limit of $\omega \to \infty$. These two extreme behaviors are mirrored in the Galilean centrifuge where one finds $\lim_{\omega \to 0} \, L = 0$ and $\lim_{\omega \to \infty} \, L = \infty$.

\subsection{Entropy}
\label{sec:entropy}

Having obtained the partition function and the ensemble averages, we can now compute the entropy \eqref{entropy_carroll_def} as:
\begin{equation}
\label{result_entropy}
S = k_B \ln \left[ \frac{4 \pi^2 m^2}{\alpha^2 \widetilde{m}^2} \left( e^{\frac{\widetilde{m}^2 R^2 \alpha}{4 q_2}} - 1 \right) \right] 
+ 2 k_B \left[ 1 - \frac{\widetilde{m}^2 R^2 \alpha}{8q_2} e^{\frac{\widetilde{m}^2 R^2 \alpha}{4q_2}} \left( e^{\frac{\widetilde{m}^2 R^2 \alpha}{4 q_2}} - 1 \right)^{-1} \right]
\end{equation}
Taking the limit $\widetilde{m}^2 \to 0$, this expression reduces to
\begin{equation}
\label{entropy_simplified}
S_{\widetilde{m}^2 = 0} = k_B \ln\left(\frac{\pi^2 m^2 R^2}{\alpha q_2}\right) +  k_B
\end{equation}

For a Galilean rotating gas, it is well-known, and obvious, that turning off rotations recovers the standard one-particle partition function without rotation. 

However, in the Carroll case, the partition function $Z$ diverges when $\omega \to 0$. Though it may seem a little strange at first glance, to some extent it is reasonable. If we only consider the time translation subgroup, the partition functions of planar Carroll systems will be divergent. Only by adding the rotation subgroup one can define the convergent partition functions. Is there a reason why thermodynamic quantities such as entropy are well-defined entirely upon the existence of rotations? A natural conjecture is that this symmetry describes systems which are forced to rotate.

Although we know nothing about the non-rotating Carroll systems, statistical physics around some small $\omega$ can still be studied. Considering $\omega$ (and hence $\alpha = \omega \beta$) very small, we have the following approximation
\begin{equation}
\label{small_omega_condition1}
e^{\frac{\widetilde{m}^2 R^2 \alpha}{4 q_2}} - 1 \approx \frac{\widetilde{m}^2 R^2 \alpha}{4q_2}, \qquad 
\text{when $|\omega| \ll \left| \frac{4 q_2}{\widetilde{m}^2 R^2 \beta} \right|$} 
\end{equation}

Under this approximation, the partition function in \eqref{partition non-zero effective mass} reduces to the same form as that in \eqref{partition zero effective mass}, \textit{i.e.} the partition function for small $\omega$ matches exactly the partition function for vanishing effective mass. The entropy is then given simply by \eqref{entropy_simplified}:
\begin{equation}
S = k_B \ln \left( \frac{\pi e m^2 A}{\omega q_2 \beta} \right) 
\end{equation}
where $A = \pi R^2$ is the area of the disc. 

Finally, we find the entropy $S \propto \ln(\kappa A)$ for some $\kappa$. Since there exist $\kappa A$ distinct occupation configurations for a planar Carroll particle, one extracts the effective area occupied by a single particle:
\begin{equation}
\label{lattice_constant_squared}
A_c = \frac{1}{\kappa} = \frac{\omega q_2 \beta}{\pi e m^2}
\end{equation}
which shows that mass and temperature act to squeeze $A_c$, while rotation and $q_2$ tend to expand $A_c$. In addition to satisfying the condition $|\omega| \ll \left| \frac{4 q_2}{\widetilde{m}^2 R^2 \beta} \right|$ imposed by \eqref{small_omega_condition1}, the angular velocity should also fulfill an additional constraint: $| \omega | \ll \left| \frac{\pi e m^2 A}{q_2 \beta}\right|$, required for the effective area $A_c$ ($\ll A $) to be meaningful.

\subsection{Equilibrium Distributions and Two-Dimensional Pressure}
\label{sec:pressure}

\subsubsection{Position Probability Density}

Thermodynamic quantities computed in the preceding subsections capture the global properties of the Carroll systems, but they tell little about how particles arrange themselves on the rotating disc. To this end, we can compute the position probability density $\rho(\bx)$ by partially integrating the phase space distribution \eqref{distribution_phase_space},
\begin{equation}
\rho(\bx) = \int_{\mathbb{R}^2} e^{\mu \cdot \xi} Z^{-1} d^2 \bm{p}
\end{equation}
Calculation is straightforward, yielding the probability density of finding a Carroll particle at radius $r$:
\begin{subequations}
\label{density_distribution}
\begin{align}
\rho(\bx) & = \rho_0 e^{\frac{\widetilde{m}^2 \omega r^2}{4 q_2 k_B T}}, \qquad r \in [0, R] \\
\rho_0  & = \frac{\widetilde{m}^2 \omega}{4 \pi q_2 k_B T} \left( e^{\frac{\widetilde{m}^2 A \omega}{4 \pi q_2 k_B T}} - 1 \right)^{-1} 
\end{align}
\end{subequations}

This profile can be compared to $\rho (\bx) \propto \exp \left( \frac{ m \omega^2 r^2}{2 k_B T} \right)$ in a Galilean centrifuge, where heavier elements accumulate farther away from the axis than lighter ones. The appearance of $\widetilde{m}^2$ for Carroll systems shows that ``heavier'' is measured by larger effective mass, rather than mass. 

\subsubsection{Velocity Distribution}

By integrating the phase space distribution over positions instead of momentum, one can obtain the velocity distribution of the system. In the Galilean case, one would obtain the Maxwell-Boltzmann law in this way. We are therefore interested in its counterpart for the Carroll gas. The velocity distribution is defined by partially integrating the phase space distribution over the spatial degrees of freedom:
\begin{equation}
f_{C}(\bm{v}) = \int_{\lvert \bm{x} \rvert \le R} 
e^{\mu \cdot \xi} Z^{-1} d^2 \bm{x}
\end{equation}
Carrying out this integral with the explicit form of the pairing \eqref{pairing_model}, we find
\begin{equation}
\label{velocity_distribution}
f_{\text{C}}(\bm{v}) = \frac{\omega q_2}{\pi k_B T} e^{- \frac{\omega q_2 v^2}{k_BT}}
\end{equation}
and thereby find the most probable speed $v_{\text{p}}$, the mean speed $v_{\text{avg}}$, and the root-mean-square speed $v_{\text{rms}}$
\begin{equation}
\label{three_speeds}
v_{\text{p}} = \sqrt{\frac{k_B T}{2 \omega q_2}}, \qquad
v_{\text{avg}} = \frac{1}{2} \sqrt{\frac{\pi k_B T}{\omega q_2}}, \qquad
v_{\text{rms}} = \sqrt{\frac{k_B T}{\omega q_2}}
\end{equation}

We see that in the Carroll case, the rotation and $q_2$ have essentially replaced the role of mass in the Maxwell-Boltzmann distribution.

\subsubsection{Pressure}

Next, we introduce the generalized Helmholtz free energy, with the standard definition:
\begin{equation}
F = U - TS - \omega L - \Phi_1 Q_1 - \Phi_2 Q_2   
\end{equation}
or, in terms of the partition function, 
\begin{equation}
F = - k_B T \ln Z
\end{equation}
Then the two-dimensional pressure\footnote{More accurately, on a two-dimensional surface $P$ is identified as a surface tension.} follows by differentiation with respect to area:
\begin{equation}
\label{def_pressure}
P = - \left( \frac{\partial F}{\partial A} \right)_{T, \, \omega, \, \Phi_1, \, \Phi_2}
= k_B T \left( \frac{\partial \ln Z}{\partial A} \right)_{T, \, \omega, \, \Phi_1, \, \Phi_2}
\end{equation}

For $N$ Carroll particles, the total partition function is $Z = Z_1^N / N!$, where $Z_1$ is the one-particle partition function given by \eqref{result_partition_function}. Using Stirling’s approximation, it is easy to find
\begin{subequations}
\begin{equation}
\ln Z \approx N \ln \left[ \frac{\pi^2 m^2 A}{N \pi \alpha q_2} 
e^{- \beta m + \zeta_1 q_1 + \zeta_2 q_2} \right] + N, \qquad \qquad \qquad \quad
\text{when $\widetilde{m}^2 = 0$}  
\end{equation}
\begin{equation}
\ln Z \approx N \ln \left[ \frac{4 \pi^2 m^2}{N \alpha^2 \widetilde{m}^2} 
e^{- \beta m + \zeta_1 q_1 + \zeta_2 q_2} \left( e^{\frac{\widetilde{m}^2 \alpha A}{4 \pi q_2}} - 1 \right) \right] + N, \qquad 
\text{when $\widetilde{m}^2 \ne 0$}
\end{equation}
\end{subequations}

We can now compute the pressure by substituting the above in \eqref{def_pressure}. In the case of zero effective mass, a direct computation recovers the ideal gas law in two-dimensional space 
\begin{equation}
P A = N k_B T
\end{equation}
and the case of non-zero effective mass gives
\begin{equation}
\label{pressure_nonzero}
P = \frac{N \widetilde{m}^2 \omega}{4 \pi q_2}  
e^{\frac{\widetilde{m}^2 \omega A}{4 \pi q_2 k_B T}} \left( 
e^{\frac{\widetilde{m}^2 A \omega}{4 \pi q_2 k_B T}} - 1 \right)^{-1}
\end{equation}
which is also the ideal gas law but on the boundary of the disc, as shown below.

Let us first look at the case of vanishing effective mass, where the position probability density $\rho$ equals a constant $\rho_0$ on the entire disc, see \eqref{density_distribution}. Thus, the number density $n = N \rho$ is also a constant, and the ideal gas law $P = n k_B T$ gives a uniform pressure everywhere. However, for the case $\widetilde{m} \ne 0$, the number density becomes non-uniform. When the number density varies with position, pressure varies accordingly as $P(r) = n(r) k_B T$. Setting $r = R$ in the equation \eqref{density_distribution}, we obtain the boundary probability density $\rho(R)$, from which we find the associated wall pressure:
\begin{equation}
\begin{split}
P(R) & = n(R) k_B T = \rho(R) N k_B T  \\
& = \frac{N \widetilde{m}^2 \omega}{4 \pi q_2}
e^{\frac{\widetilde{m}^2 \omega A}{4 \pi q_2 k_B T}} \left( 
e^{\frac{\widetilde{m}^2 A \omega}{4 \pi q_2 k_B T}} - 1 \right)^{-1}
\end{split}
\end{equation}
This means that the gas law \eqref{pressure_nonzero} is precisely the ideal gas law, but evaluated on the boundary of the disc.

\subsection{Effective Hamiltonian}

The argument in Section \ref{sec:statistical_physics} illustrates two ways to construct the partition function. So far, we have considered the second approach that introduces a non-zero angular momentum rather than the notion of centrifugal potential. This brings us to a natural question: what is the effective Hamiltonian of a rotating Carroll system viewed from the perspective of the first approach? Hence, what we shall do next is derive it via the variable transformation $\bm{p}_{\text{rotating}} = \bm{p}_{\text{inertial}} + \bm{p}_{\text{rel}}$ between the rotating and inertial frames.

Recall that the momentum transformation in $3+1$ dimensional Galilean theory is
\begin{equation}
\bm{p}_{\text{rotating}} = \bm{p}_{\text{inertial}} - m \bm{\omega} \times \bm{x}
\end{equation}
This, however, is not true in Carroll theory, because there the relative momentum $\bm{p_{\text{rel}}} = m \bm{v_{\text{rel}}}$ is completely decoupled from the time derivative of position $d \bm{x} / dt = - \bm{\omega} \times \bm{x}$. 

Let the origins of the two reference frames coincide. Locally (in the neighborhood of a fixed point) and instantaneously (for an infinitesimal time interval), the relative factor $\bm{v}_{\text{rel}}$ behaves as a constant vector, corresponding to a shift $\bm{v} \mapsto \bm{v} + \bm{v}_{\text{rel}}$. Consequently, $\bm{v}_{\text{rel}}$ is equivalent to a boost\footnote{We emphasize that it does not possess a global Carroll boost symmetry. Nevertheless, for a chosen point $\bm{x}$ and the instantaneous limit $\Delta t \to 0$ we consider here, equation \eqref{pre_proper_trans} always holds.} associated with which one can use the conserved quantity \eqref{conserved_quantity_boost} of the extended Carroll group:
\begin{equation}
\label{pre_proper_trans}
0 = \delta \bm{g} 
= m \delta \bm{x} + \frac{2q_2}{m} \epsilon \delta \bm{p}
\end{equation}
the above equation implies the existence of a constant $C$, such that
\begin{equation}
\label{relative_momentum}
C = m \bm{x} + \frac{2q_2}{m} \epsilon \bm{p_{\text{rel}}}   
\end{equation}

We set the constant to zero because we want the origin to have the same momentum in both the rotating and the inertial systems, \textit{i.e.} if $\bm{x} = \bm{0}$, then $\bm{p_{\text{rel}}} = \bm{0}$. This condition uniquely determines the solution of \eqref{relative_momentum} and thus gives the value of the relative momentum as $\bm{p}_{\text{rel}} = m^2 \epsilon \bm{x} / 2q_2$.

The proper transformation for any point $\bm{x}$ is therefore given by
\begin{equation}
\bm{p}_{\text{rotating}} = \bm{p}_{\text{inertial}} + \frac{m^2}{2q_2} \epsilon \bm{x} 
\end{equation}
Comparison of the relative momentum, $m \omega \epsilon \bm{x}$ for the Galilean case and $\frac{m^2}{2q_2} \epsilon \bm{x}$ for the Carroll case, suggests that $q_2$ is related to some intrinsic rotation. Inserting this (canonical) transformation $p_1 \mapsto p_1 + \frac{m^2}{2q_2} x_2$ \& $p_2 \mapsto p_2 - \frac{m^2}{2q_2} x_1$ into the partition function
\begin{equation}
Z = e^{\zeta_1 q_1 + \zeta_2 q_2} 
\int_{\cU} e^{- \beta m + \alpha \left(\bm{p} \times \bm{x} + q_1 \bm{x}^2 - \frac{q_2}{m^2} \bm{p}^2 \right)} 
d^2 \bm{p} d^2 \bm{x} 
\end{equation}
we find the new, but equivalent, expression
\begin{equation}
Z = e^{\zeta_1 q_1 + \zeta_2 q_2} 
\int_{\cU} e^{- \beta \left( \frac{\omega q_2}{m^2} \bm{p}^2 - \frac{\widetilde{m}^2 \omega}{4 q_2} \bm{x}^2 + m \right)} 
d^2 \bm{p} d^2 \bm{x}
\end{equation}

Identifying the factor in the exponent of the above formula with the structure $- \beta H$, one recasts the effective Hamiltonian
\begin{equation}
\label{effective_Hamiltonian}
H(\bm{p}, \bm{x}) = 
- \frac{\widetilde{m}^2 \omega}{4 q_2} \bm{x}^2 
+ \frac{\omega q_2}{m^2} \bm{p}^2 
+ m
\end{equation}
The first component is the centrifugal potential in Carroll theory, and interestingly, the second component is akin to a kinetic term that vanishes in non-rotating and non centrally extended Carroll dynamics. Similarly to the velocity distribution of the previous subsection, we note that the rotation and $q_2$ play the role of mass.

\section{Discussion}
\label{sec:conclusions}

In this article, we established a well-defined notion of statistical physics for planar Carroll systems. Applying Souriau's geometric thermodynamics, we showed that (at least one of) the two central extension charges of the planar Carroll algebra must be included as physical parameters in the partition function to ensure convergence.

It was shown in \cite{deBoer:2023fnj} that the partition functions of Carroll systems obtained as the zero speed of light limit of Poincaré ones tend to diverge and do not lead to well-behaved thermodynamics. As the authors stress, this appears to be an inherent property of the Carroll system, because the energy $E$ associated with time translations does not depend on momentum at all, causing the partition function to diverge when one integrates $e^{-\beta E}$ over the momentum. Here, we showed that considering angular momentum \eqref{conserved_quantity_rotations}, which is the only physical parameter containing the square of the momentum, together with one of the central extensions, allows the partition functions to converge.

Hence, we considered a rotating disc characterized by angular velocity $\omega$ and investigated co‑moving Carroll particles on it. The first law of thermodynamics \eqref{first_law} allowed us to calculate the pressure of this rotating gas, and the computation in Section \ref{sec:pressure} shows that it obeys the standard ideal gas law in $2 + 1$ dimensions. Furthermore, we derived the partition function \eqref{result_partition_function}, the average angular momentum \eqref{result_average_angular_momentum}, and the entropy \eqref{result_entropy} for a single Carroll particle. 

Crucially, all of these quantities converge under the condition \eqref{convergence_condition}:
\begin{equation*}
\omega q_2 > 0
\end{equation*} 
The above equilibrium condition suggests that Carroll systems with positive $q_2$ can reach thermal equilibrium only if they rotate clockwise, while those systems with negative $q_2$ must rotate counterclockwise. Remarkably, rotation is inevitable because this condition forbids a vanishing angular velocity.

Given the link between rotations and $q_2$, we wonder if this charge can be related to some system with intrinsic rotations. One potential link could be given in condensed matter, where vortices have been shown to realize fractons in some cases \cite{DoshiG20}, and as we explained in the introduction, fractons seem to be realized with the Carroll algebra.

Every thermodynamic quantity we have computed involves a key parameter: the effective mass squared $\widetilde{m}^2 = m^2 + 4 q_1 q_2$, which includes, in particular, the two central extension charges. This effective mass already appears in previous studies on the dynamic of planar particles \cite{Marsot21,Marsot:2022imf}. Here, it provides a measure of ``weight'', replacing the role of mass in the Galilean centrifuge. This result marks that the position probability density \eqref{density_distribution} of the rotating Carroll particles is similar to the Galilean case, in that particles with larger $\widetilde{m}^2$ are statistically found further apart from the axis of rotation. We also obtained the corresponding velocity distribution \eqref{velocity_distribution} and the three characteristic speeds \eqref{three_speeds}, in parallel with the classical Maxwell–Boltzmann distribution.

When studying the small $\omega$ approximation or vanishing effective mass in Section \ref{sec:entropy}, the entropy is greatly simplified and ultimately yields $S \propto \ln(\kappa A)$, with $\kappa$ a constant and $A$ the area of the system. Although this entropy behavior can be expected in the plane, this is interesting given the link between Carroll symmetry and black hole horizon, specifically that the horizon is a Carroll geometry. Although the famous Bekenstein-Hawking black hole entropy depends on the area of the black hole, there has been many studies showing that one expects a logarithmic correction to this entropy, see \textit{e.g.} \cite{Kaul:2000kf,Carlip:2000nv,Gour:2003jj,PhysRevD.71.027502,Medved:2004yu,Sen:2012dw}, and it could be interesting to study whether these quantum corrections responsible for the logarithmic contribution could be realized as some sort of Carrollian system, as this would provide a physical interpretation for $q_2$.

A recent paper \cite{kinetictheorycarrollhydrodynamics} also establishes a notion of Carroll statistical physics from first principles, however in a different way. The authors employ the distribution function of space-filling branes $t(\bm{x})$ which replaces particle's trajectory $\bm{x}(t)$, motivated by the exchange of the roles of time and space between Carroll and Galilean physics. They further derive Carrollian fluid equations from brane collisions, while our study concentrates on under what conditions a Carroll system can reach equilibrium, as well as the resulting Carrollian equilibrium thermodynamics.

\section*{Acknowledgements}

The authors are grateful to Salvatore Ribisi, Francesco Toppan and Zhanna Kuznetsova for useful discussions. This work is supported by the National Natural Science Foundation of China (Grant No. W2433003).

\end{document}